\documentclass[aps,prl,twocolumn,showpacs,groupedaddress]{revtex4}  % for review and submission 

\usepackage{graphicx}% Include figure files
\usepackage{dcolumn}% Align table columns on decimal point
\usepackage{bm}% bold math

\newcommand{\ptmax}{p_T^{\text{max}}}
\newcommand{\Mjj}{M_{\text{\sl jj}}}
\newcommand{\chijj}{\chi_{\text{dijet}}}

\newcommand{\yboost}{y_{\text{boost}}}

\newcommand{\pythia}{{\sc pythia}}
\newcommand{\nlojet}{{\sc nlojet++}}

\newcommand{\ppbar}{p{\bar{p}}}

\begin{document}
\hspace{5.2in} \mbox{FERMILAB-PUB-09-326-E}

\title{\boldmath Measurement of Dijet Angular Distributions
at $\sqrt{s}=1.96\,$ TeV and Searches \\
for Quark Compositeness and Extra Spatial Dimensions}

% LIST_OF_AUTHORS_R2.TEX                 5/15/09            
%
\author{V.M.~Abazov$^{37}$}
\author{B.~Abbott$^{75}$}
\author{M.~Abolins$^{65}$}
\author{B.S.~Acharya$^{30}$}
\author{M.~Adams$^{51}$}
\author{T.~Adams$^{49}$}
\author{E.~Aguilo$^{6}$}
\author{M.~Ahsan$^{59}$}
\author{G.D.~Alexeev$^{37}$}
\author{G.~Alkhazov$^{41}$}
% ...

\affiliation{\vspace{0.1 in}(The D\O\ Collaboration)\vspace{0.1 in}}
% ...
\affiliation{$^{6}$University of Alberta, Edmonton, Alberta, Canada;
                Simon Fraser University, Burnaby, British Columbia, Canada;
                York University, Toronto, Ontario, Canada and
                McGill University, Montreal, Quebec, Canada}
% ...
\affiliation{$^{30}$Tata Institute of Fundamental Research, Mumbai, India}
% ...
\affiliation{$^{37}$Joint Institute for Nuclear Research, Dubna, Russia}
% ...
\affiliation{$^{41}$Petersburg Nuclear Physics Institute,
                St. Petersburg, Russia}
% ...
\affiliation{$^{49}$Florida State University, Tallahassee, Florida 32306, USA}
% ...
\affiliation{$^{51}$University of Illinois at Chicago,
                Chicago, Illinois 60607, USA}
% ...
\affiliation{$^{59}$Kansas State University, Manhattan, Kansas 66506, USA}
% ...
\affiliation{$^{65}$Michigan State University,
                East Lansing, Michigan 48824, USA}
% ...
\affiliation{$^{75}$University of Oklahoma, Norman, Oklahoma 73019, USA}
% ...

\date{Received 29 June 2009; published 5 November 2009}
           
\begin{abstract}
We present the first measurement of dijet angular distributions
in $\ppbar$ collisions at $\sqrt{s}=1.96\,$TeV 
at the Fermilab Tevatron Collider.
The measurement is based on a dataset corresponding to an 
integrated luminosity of $0.7\,$fb$^{-1}$ 
collected with the D0 detector.
Dijet angular distributions have been measured over a range 
of dijet masses, from $0.25\,$TeV to above $1.1\,$TeV.
The data are in good agreement with the predictions of 
perturbative QCD and are used to constrain  
new physics models including quark compositeness,
large extra dimensions, and TeV$^{-1}$ scale extra dimensions.
For all models considered, we set the most stringent direct limits 
to date.
\end{abstract}

\pacs{12.60.Rc, 11.25.Wx, 12.38.Qk, 13.87.Ce}

\maketitle

% ************************************************************************
% *************          Introduction
% ************************************************************************

At large momentum transfers, dijet production has the largest cross section 
of all processes at a hadron collider and therefore directly probes 
the highest energy regime.
It can be used to test the standard model (SM) at previously unexplored 
small distance scales and to search for signals predicted 
by new physics models.
The angular distribution of dijets with respect to the 
hadron beam direction is directly sensitive to the 
dynamics of the underlying reaction.
While in quantum chromodynamics (QCD) this distribution 
shows small but noticeable deviations from Rutherford scattering,
an excess at large angles from the beam axis
would be a sign of new physics processes not included in the SM, 
such as substructure of quarks 
(``quark compositeness'')~\cite{Eichten:1984eu,Chiappetta:1990jd,Lane:1996gr}, 
or the existence of additional compactified spatial dimensions
(``extra dimensions'')~\cite{ArkaniHamed:1998rs,Atwood:1999qd,Dienes:1998vg,Pomarol:1998sd,Cheung:2001mq}.
Earlier measurements of dijet angular distributions 
and related observables in $\ppbar$ collisions at $\sqrt{s}=1.8\,$TeV
were used to set limits on 
quark compositeness~\cite{Abe:1996mj,Abbott:2000kp}.

In this Letter we present the first measurement of dijet angular
distributions in $p\bar{p}$ collisions at a center-of-mass 
energy of $\sqrt{s}=1.96$\,TeV.
The data sample, collected with the D0 detector during 2004--2005
in Run~II of the Fermilab Tevatron Collider, corresponds to
an integrated luminosity of $0.7\,$fb$^{-1}$.
In the experiment and in theory calculations,
jets are defined by the Run~II midpoint cone jet algorithm~\cite{run2cone} 
with a cone radius of 
${\cal R}=\sqrt{(\Delta y)^2 + (\Delta \phi)^2}=0.7$ in rapidity $y$ 
and azimuthal angle $\phi$.
Rapidity is related to the polar scattering angle $\theta$
with respect to the beam axis
by $y=0.5 \ln \left[ (1+\beta \cos \theta) / (1-\beta \cos \theta) \right]$
with $\beta=|\vec{p}| / E$.
We measure distributions in the dijet variable
$\chijj = \exp(|y_1 - y_2|)$
in ten regions of dijet invariant mass $\Mjj$, 
where $y_1$ and $y_2$ are the rapidities of the two jets with
highest transverse momentum $p_T$ with respect to the beam axis
in an event.
For massless $2\rightarrow 2$ scattering, the
variable $\chijj$ is related to the polar scattering angle
$\theta^*$ in the partonic center-of-mass frame by
$\chijj = (1+\cos\theta^*)/(1-\cos\theta^*)$.
The choice of this variable is motivated by the fact 
that Rutherford scattering is independent of $\chijj$.
The phase space of this analysis is defined by
$\Mjj>0.25\,$TeV, $\chijj < 16$, and 
$\yboost = 0.5 \, |y_1 + y_2|  <1$.
Together, the $\chijj$ and $\yboost$ requirements restrict 
the jet phase space to $|y_{\text{jet}}|<2.4$
where jets are well-reconstructed in the D0 detector 
and the energy calibration is known to high precision.
To minimize sensitivity to correlated experimental and theoretical 
uncertainties, 
the $\chijj$ distributions in the different $\Mjj$ ranges 
are normalized by their respective integrals.
Based on the measurement, we set limits on
quark compositeness~\cite{Eichten:1984eu,Chiappetta:1990jd,Lane:1996gr}, 
large spatial extra dimensions according to the model proposed by
Arkani-Hamed, Dimopoulos and Dvali
(ADD LED)~\cite{ArkaniHamed:1998rs,Atwood:1999qd},
and TeV$^{-1}$ scale extra dimensions 
(TeV$^{-1}$ ED)~\cite{Dienes:1998vg,Pomarol:1998sd,Cheung:2001mq}.

% *********************************************************************
% ***************   Measurement
% *********************************************************************

A detailed description of the D0 detector can be found in
Ref.~\cite{d0det}.
The event selection, jet reconstruction, jet energy and 
momentum correction in this measurement follow closely 
those used in our recent measurement of the inclusive 
jet cross section~\cite{:2008hua}.
The primary tool for jet detection is the
finely segmented uranium-liquid argon calorimeter that
has almost complete solid angular coverage 
$1.7^\circ \lesssim \theta \lesssim 178.3^\circ$~\cite{d0det}.
Events are triggered by the jet with highest $p_T$, referred to as $\ptmax$.
In each $\Mjj$ region, events are taken from a single trigger
which is chosen such that the smallest $\ptmax$ in the $\Mjj$ region 
is above the threshold that ensures 100\% efficiency.
The $\Mjj$ regions utilize triggers with different prescales, 
resulting in integrated luminosities of
0.10\,pb$^{-1}$ ($\Mjj<0.4\,$TeV), 
1.54\,pb$^{-1}$  ($0.4<\Mjj<0.5\,$TeV), 
17\,pb$^{-1}$ ($0.5<\Mjj<0.6\,$TeV), 
73\,pb$^{-1}$ ($0.6<\Mjj<0.8\,$TeV), 
0.5\,fb$^{-1}$ ($0.8<\Mjj<1.0\,$TeV), 
and 0.7\,fb$^{-1}$  ($\Mjj>1.0\,$TeV).

% ********************************************************************
% **************   Jet Correction: JES, vertex, MET, jet ID
% ********************************************************************

The position of the $p\bar{p}$ interaction is reconstructed using a
tracking system consisting of silicon microstrip detectors and 
scintillating fibers, located inside a $2\,\text{T}$ solenoidal 
magnet~\cite{d0det}, 
and is required to be within $50$\,cm of the detector center
along the beam direction. 
The jet four-momenta are corrected for the response of the calorimeter, 
the net energy flow through the jet cone,
energy from event pile-up and multiple $p\bar{p}$ interactions,
and for systematic shifts in $|y|$ due to detector effects~\cite{:2008hua}.
Cosmic ray backgrounds are suppressed by requirements on 
the missing transverse momentum in an event~\cite{:2008hua}.
Requirements on characteristics of shower shape are
used to suppress the remaining background due to electrons, photons, 
and detector noise that mimic jets. 
The efficiency for these requirements is above $97.5\%$, 
and the fraction of background events is below $0.1$\% in all $\Mjj$ regions.

% ---------------------------------------------------------
% -----------     correction, simulation
% ---------------------------------------------------------

The $\chijj$ distributions are corrected for instrumental effects
using events generated with \pythia\/ v6.419~\cite{pythia} 
using tune QW~\cite{Albrow:2006rt}
and MSTW2008LO parton distribution functions (PDFs)~\cite{Martin:2009iq}.
The generated particle-level events are subjected to a fast simulation of the
D0 detector response, based on parametrizations of 
resolution effects in $p_T$, the polar and azimuthal angles of jets,
jet reconstruction efficiencies, and misidentification of the 
event vertex.
These parametrizations have been determined either from data or 
from a detailed simulation of the D0 detector using {\sc geant}~\cite{geant}.
The generated events are reweighted according to the $\Mjj$ distribution
in data.
To minimize migrations between $\Mjj$ regions
due to resolution effects,
we use the simulation to obtain a rescaling function in $\Mjj$
that optimizes the correlation between the reconstructed and true values.
The bin sizes in the $\chijj$ distributions are chosen to be much 
larger than the $\chijj$ resolution.
The bin purity after $\Mjj$ rescaling, defined as the fraction
of all reconstructed events that were generated in the same
bin, is between $42\%$ and $68\%$.
We then use the simulation to determine $\chijj$ bin correction factors
for the differential cross sections in the different $\Mjj$ regions.
These also include corrections for the energies of unreconstructed
muons and neutrinos inside the jets.
The total correction factors for the differential cross sections are 
typically between 0.9 and 1.0, and always in the range 0.7 to 1.1.
The corrected differential cross sections within each $\Mjj$ range
are subsequently normalized to their integrals, providing
the corrected, final results for 
$1/\sigma_{\text{dijet}}\cdot d\sigma / d\chijj$
at the ``particle level'' as defined in Ref.~\cite{Buttar:2008jx}.

In order to take into account correlations between systematic uncertainties,
the experimental systematic uncertainties are separated into independent 
sources, for each of which the effects are fully correlated between all 
data points.
In total we have identified 76 independent sources, 
of which 48 are related to the jet energy calibration
and 15 to the jet $p_T$ resolution uncertainty. 
These are the dominant sources of uncertainty.
Smaller contributions are from the jet $\theta$ resolution and from 
the systematic shifts in $y$.
All other sources are negligible.
All sources and their effects are documented in Ref.~\cite{chidata}.
For $\Mjj<1$\,TeV ($\Mjj>1$\,TeV) systematic uncertainties are
1\%--5\% (3\%--11\%);
they are in all cases less than the statistical uncertainties.

\begin{figure}
\includegraphics[scale=0.8]{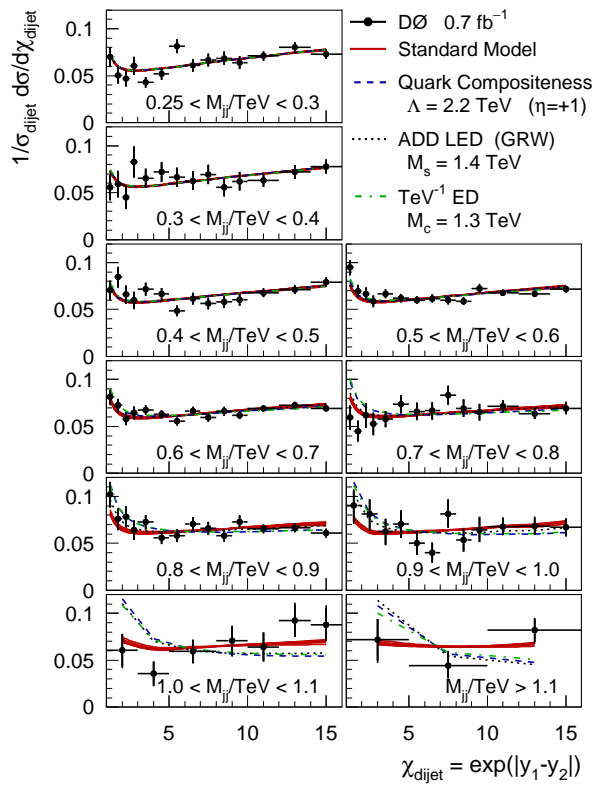}% 
\caption{\label{fig:result}  
  Normalized differential cross sections in $\chijj$ compared to
  standard model predictions and to the predictions of various
  new physics models.
  The error bars display the quadratic sum 
  of statistical and systematic uncertainties.
  The standard model theory band includes uncertainties from scale variations
  and PDF uncertainties (see text for details).}
\end{figure}

The results are available in Ref.~\cite{chidata} 
and displayed in Fig.~\ref{fig:result}.
The normalized $\chijj$ distributions are presented in ten $\Mjj$ regions, 
starting from $\Mjj>0.25\,$TeV, and including one region for $\Mjj>1.1\,$TeV.
The data are compared to 
predictions from a perturbative 
QCD calculation at next-to-leading order (NLO)
with non-perturbative corrections applied.
The non-perturbative corrections are determined using \pythia.  
They are defined as the product of the corrections due to hadronization 
and to the underlying event.
The NLO results are computed using {\sc fastnlo}~\cite{Kluge:2006xs}
based on \nlojet~\cite{Nagy:2003tz,Nagy:2001fj}.
All theory calculations use MSTW2008NLO PDFs~\cite{Martin:2009iq} and 
the corresponding value of $\alpha_s(M_Z)=0.120$.
The PDF uncertainties are provided by the twenty 
MSTW2008NLO 90\% C.L.\ eigenvectors.
Renormalization and factorization scales $\mu$ are varied simultaneously
around the central value of $\mu_0 = \langle p_T\rangle$ in the range
$0.5\, \mu_0 \le \mu \le 2\,\mu_0$
where $\langle p_T\rangle$ is the average dijet $p_T$.
The quadratic sum of scale and PDF uncertainties is displayed 
as a band around the central SM value in Fig.~\ref{fig:result}.
The scale (PDF) uncertainties are always below 
5\% (2\%) so the band is nearly a line.
The theory, including the perturbative results
and the non-perturbative corrections, is in good agreement with the data
over the whole $\Mjj$ range with a $\chi^2$ (defined later) of $127.2$
for 120 data points in ten normalized distributions.
Based on the agreement of the $\chijj$ measurement with the SM,
we proceed to set limits on quark compositeness, 
ADD LED, and TeV$^{-1}$ ED models.

% *********************************************************************
% ************  Interpretation
% *********************************************************************

Hypothetically, quarks could be made of other particles,
as assumed in the quark compositeness 
model in Ref.~\cite{Eichten:1984eu,Chiappetta:1990jd,Lane:1996gr}.  
We investigate the model in which all quarks are considered
to be composite. 
The parameters in this model are the energy scale $\Lambda$ 
and the sign of the interference term $\eta$ between the standard 
model and the new physics terms.
The ADD LED model~\cite{ArkaniHamed:1998rs,Atwood:1999qd} assumes that
extra spatial dimensions exist in which gravity is allowed to propagate.
Jet cross sections receive additional contributions from
virtual exchange of Kaluza-Klein excitations of the graviton.
There are two different formalisms (GRW~\cite{Giudice:1998ck} 
and HLZ~\cite{Han:1998sg}).
The model parameter is the effective Planck scale, $M_S$, and
the HLZ formalism also includes the subleading 
dependence on the number $n$ of extra dimensions.
The  TeV$^{-1}$ ED model~\cite{Dienes:1998vg,Pomarol:1998sd,Cheung:2001mq} 
assumes that extra dimensions exist at the TeV$^{-1}$ scale.
SM production cross sections are modified due to virtual 
Kaluza-Klein excitations of the SM gauge bosons.
In this model, gluons can travel through the extra dimensions,
which changes the dijet cross section.
The parameter in this model is the compactification scale, $M_C$.

\begin{table*}
\caption{Expected and observed 95\% C.L.\ limits in units of TeV 
on various new physics (NP) models for different Bayesian priors,
and for the $\Delta \chi^2$ criterion.
\label{tab:limits}}
\begin{ruledtabular}
\begin{tabular}{l c c c c c c}
  & \multicolumn{2}{l}{Prior flat in NP Lagrangian}
  & \multicolumn{2}{l}{Prior flat in NP $x$-section} 
 & \multicolumn{2}{c}{$\Delta \chi^2=3.84$ criterion}  \\
Model (parameter)  & \multicolumn{1}{c}{Expected} 
  & \multicolumn{1}{c}{Observed} & 
  \multicolumn{1}{c}{Expexted}   & \multicolumn{1}{c}{Observed}   &
  \multicolumn{1}{c}{Expected}   & \multicolumn{1}{c}{Observed}   \\ 
\hline
Quark compositeness ($\Lambda$) & & & & \\
\phantom{m}   $\eta=+1$      &   
     2.91  &  3.06 &   
     2.76  &  2.84 &   
     2.80  & 2.92   \\
\phantom{m}   $\eta=-1$      &   
     2.97  &   3.06 &    
     2.75  &   2.82 &    
     2.82 & 2.96 \\
TeV$^{-1}$ ED ($M_C$)  &
     1.73  & 1.67 &     
     1.60  & 1.55 &     
     1.66 & 1.59   \\
ADD LED ($M_S$) &  & & &   \\
\phantom{m} GRW &   
     1.53  & 1.67 &   
     1.47  & 1.59 &   
     1.49 & 1.66  \\
\phantom{m} HLZ $\; n=3$ & 
     1.81  & 1.98 &   
     1.75  & 1.89 &    
     1.77 & 1.97  \\
\phantom{m} HLZ $\; n=4$ & 
     1.53  & 1.67 &   
     1.47  & 1.59 &    
    1.49 & 1.66 \\
\phantom{m} HLZ $\; n=5$ & 
     1.38  & 1.51 &  
     1.33  & 1.43 &  
     1.35 & 1.50 \\
\phantom{m} HLZ $\; n=6$ & 
     1.28  & 1.40 &  
     1.24  & 1.34 &   
     1.25 & 1.39 \\
\phantom{m} HLZ  $\; n=7$ & 
     1.21  & 1.33 &   
     1.17  & 1.26 &    
    1.19 & 1.32 \\
\end{tabular}
\end{ruledtabular}
\end{table*}

The new physics contributions have only been calculated
to leading order (LO), while the QCD predictions are known to NLO.
In this analysis, to obtain the best estimate for 
new physics processes, 
we multiply the new physics LO calculations bin-by-bin
by the SM $k$-factors ($k=\sigma_{\text{NLO}}/\sigma_{\text{LO}}$).
The $k$-factors are in the range 1.25--1.5, increasing with $\Mjj$ 
and decreasing with $\chijj$.
Their effects on single bins of the normalized $\chijj$ distributions 
within the different $\Mjj$ regions is below 12\%.
The new physics cross sections are computed using the matrix elements from 
Refs.~\cite{Chiappetta:1990jd,Lane:1996gr,Atwood:1999qd,Cheung:2001mq}.
The theoretical variations (scale variations
and PDF uncertainties) are
consistently propagated into both the SM and the new physics contributions.
Predictions for the different models are compared to the
$\chijj$ data and to the SM results in Fig.~\ref{fig:result}.
It is observed that all models predict increased contributions as
$\chijj \rightarrow 1$ towards large $\Mjj$.
The $\Mjj$ evolution of the excess towards small $\chijj$ is observed
to be different for different models.

% --------------------------------------------------------------
% ----------- Limits
% --------------------------------------------------------------

We define the $\chi^2$ between data and theory using the 
Hessian approach~\cite{Alekhin:2005dx} which introduces 
nuisance parameters for all correlated sources of
experimental and theoretical uncertainty.
The $\chi^2$ is then minimized with respect to all nuisance 
parameters, and is therefore only a function of the new physics 
model parameter(s).   
In most cases $\chi^2$ has the minimum for a new physics mass scale 
of infinity, corresponding to the SM value.
Only for the quark compositeness model with positive interference
and for the TeV$^{-1}$ ED model $\chi^2$ has small minima
at $\Lambda=9.88\,$TeV with $\Delta\chi^2=0.01$
and $M_C=2.96\,$TeV with $\Delta\chi^2=0.28$
below the SM value, respectively.

The $\chi^2$ is then transformed into a likelihood which is used in a 
Bayesian procedure~\cite{Abbott:2000kp} to obtain 95\% C.L. limits on
the new physics mass scales $\Lambda$, $M_C$, and $M_S$
in the different models.
The prior is chosen to be flat in the new physics mass scale 
raised to the power in which it appears in the Lagrangian or, 
alternatively, 
raised to the power in which it enters the model cross section.
While the former has been used in many previous analyses,
the latter is statistically preferred for being unbiased 
in the cross section.
Alternatively, we have applied a procedure which defines the 
95\% C.L. limit as the mass scale at which 
$\chi^2 - \chi^2_{\text{min}} = 3.84$~\cite{pdg}.
This procedure has the advantage of being independent 
of an assumed prior.
The observed limits and the expectation values 
are listed in Table~\ref{tab:limits}.
All observed limits are within one standard deviation of the expected limits.

The limit on $M_C$ obtained in this analysis, 
while inferior to indirect limits from electroweak precision 
measurements (Ref.~\cite{Cheung:2001mq} and references therein),
is complementary and is the result of the first direct
search for TeV$^{-1}$ extra dimensions at a particle collider.
The limits on $M_S$ in the different formalisms of the ADD LED model
are on average slightly higher as compared to recent D0 results from 
the combination of 1\,fb$^{-1}$ of dielectron and diphoton data 
in Ref.~\cite{Abazov:2008as}, 
which were so far the most restrictive limits on ADD LED.
Our limits on quark compositeness improve previous results 
from related dijet observables~\cite{Abbott:2000kp,Abe:1996mj}
and are the most stringent limits to date.

% --------------------------------------------------------------
% -----------  Summary
% --------------------------------------------------------------
In summary, we have presented the first measurement of dijet angular 
distributions in Run II of the Fermilab Tevatron Collider.
This is the first measurement of angular distributions 
of a hard partonic scattering process at energies
above 1\,TeV in collider-based high energy physics.
The normalized  $\chijj$ distributions are well-described by
theory calculations in next-to-leading order in the strong coupling constant
and are used to set limits on quark compositeness, ADD large extra dimensions, 
and TeV$^{-1}$ extra dimensions models.
For the TeV$^{-1}$ extra dimensions model this is the first
direct search at a collider.
For all models considered, this analysis sets the most 
stringent direct limits to date.

\begin{acknowledgments}
% acknowledgement_paragraph_r2.tex                         5/15/09
%
We thank the staffs at Fermilab and collaborating institutions, 
and acknowledge support from the 
DOE and NSF (USA);
CEA and CNRS/IN2P3 (France);
FASI, Rosatom and RFBR (Russia);
CNPq, FAPERJ, FAPESP and FUNDUNESP (Brazil);
DAE and DST (India);
Colciencias (Colombia);
CONACyT (Mexico);
KRF and KOSEF (Korea);
CONICET and UBACyT (Argentina);
FOM (The Netherlands);
STFC and the Royal Society (United Kingdom);
MSMT and GACR (Czech Republic);
CRC Program, CFI, NSERC and WestGrid Project (Canada);
BMBF and DFG (Germany);
SFI (Ireland);
The Swedish Research Council (Sweden);
CAS and CNSF (China);
and the
Alexander von Humboldt Foundation (Germany).

\end{acknowledgments}

\end{document}